\documentclass[a4paper,UKenglish,cleveref, autoref, thm-restate]{oasics-v2021}

\usepackage{todonotes}
\usepackage{comment}

\title{On Performance Analysis of Graphcore IPUs: Analyzing Squared and Skewed Matrix Multiplication}

\author{S.-Kazem Shekofteh}{Institute of Computer Engineering, Heidelberg University, Germany}{kazem.shekofteh@ziti.uni-heidelberg.de}{https://orcid.org/0000-0002-8783-6243}{}

\author{Christian Alles}{Institute of Computer Engineering, Heidelberg University, Germany}{christian.alles@stud.uni-heidelberg.de}{[orcid]}{}

\author{Nils Kochendörfer}{Institute of Computer Engineering, Heidelberg University, Germany}{nils.kochendoerfer@stud.uni-heidelberg.de}{[orcid]}{}

\author{Holger Fröning}{Institute of Computer Engineering, Heidelberg University, Germany}{holger.froening@ziti.uni-heidelberg.de}{https://orcid.org/0000-0001-9562-0680}{}

\authorrunning{S.-Kazem  Shekofteh} 

\Copyright{S.-Kazem  Shekofteh} 

\ccsdesc[500]{Hardware~Analysis and design of emerging devices and systems}
\ccsdesc[500]{Computing methodologies~Massively parallel algorithms}

\keywords{matrix multiply, parallel processing, intelligence processing units} 

\category{} 

\relatedversion{} 

\nolinenumbers 

\EventShortTitle{preprint}
\EventAcronym{preprint}
\EventYear{2023}
\EventDate{January 17, 2023}
\EventLocation{?}
\EventLogo{}

\begin{document}

\maketitle

\begin{abstract}
In recent decades, High Performance Computing (HPC) has undergone significant enhancements, particularly in the realm of hardware platforms, aimed at delivering increased processing power while keeping power consumption within reasonable limits.
The Intelligence Processing Unit (IPU) represents an entirely novel category of massively parallel processors meticulously designed to expedite parallel computations through a multitude of processing cores and on-chip memory components interconnected via high-speed fabrics. While IPUs are primarily tailored for machine learning applications and come equipped with several libraries for the seamless implementation of neural networks, they also retain the capability to execute traditional parallel programs like matrix multiplication. However, it is essential to acknowledge that there are certain considerations and limitations when utilizing IPUs for such tasks.
This paper embarks on an extensive analytical examination of matrix multiplications (MM) executed on an IPU, focusing on aspects such as execution efficiency and memory usage. Additionally, a comparative analysis is conducted, pitting the IPU against a GPU. Our findings indicate that IPUs can outperform modern GPUs, especially in handling the consistently challenging skewed matrix multiplication operations. For a more comprehensive understanding, we scrutinize various aspect ratios of matrices for these operations on an IPU and a Turing-class GPU (RTX 2080TI), revealing that the IPU consistently delivers more robust performance when dealing with skewed matrices compared to a GPU.
\end{abstract}

\section{Introduction}
\label{sec:typesetting-summary}

During last decade, Graphics Processing Units (GPUs) became ubiquitous as \textit{accelerators} for performance improvement in almost all sciences due to their computaional power, ease-of-use capabilities for programming, strong support by manufacturers and the community to develop new libraries and advance the stir to benefit from these accelerators.
Contrary, the Intelligence Processing Unit (IPU) is a completely new type of accelerator, which is proposed to satisfy the increasing demand for hardware resources in Machine Learning (ML) applications such as neural networks. 
IPU processors are optimized to perform highly-parallel fine-grained operations. 

In contrast to the single instruction multiple threads (SIMT) architecture of GPUs, which requires contiguous vectorized data for efficient operation, the IPU is highly efficient on applications with both \textit{dense and irregular} sparse data access, and can run individual processing threads on small data blocks while exploiting its multiple instruction multiple data (MIMD) architecture \cite{ipu1}.

Most literature dealing with the IPU uses this processor to accelerate ML applications with frameworks such as TensorFlow and PyTorch. 
A common insight is the limited amount of memory and the high computational performance potential. Specifically, \cite{ipu_ML_bench1}, \cite{Dyn_Sparse}, and \cite{Qwant} try to exploit the IPUs for running and evaluating machine learning algorithms, so that they can benefit from the maximum advertised perforamnce of the IPU. More precisely,   \cite{ipu_ML_bench1} talks about key differences between GPUs and IPUs when running AI algorithms in neural network training and inference.

A prime example for dissecting the performance of a given hardware is the matrix multiply (MM) operation, as such calculations are often used in various scientific fields  \cite{GEMM}. In other work, these were restricted to MM of squared matrices by using library functions. A further look into MM of skewed matrices is therefore necessary. Since no mention was made how scarce memory is used in previous work, investigation in that area also has to be made.

Thus, we focus in this work on non-square MM operations, and gear to provide an analysis of performance of such operations on the IPU in comparison to a recent GPU as a baseline. In this regard, the present work makes the following contributions:

\begin{enumerate}
\item First, we compare the overall performance or standard, i.e., squared, MM on IPU and GPU using provided libraries to the theoretical peak performance of these processors. 
\item Second, a comparison for skewed MM is presented between GPU and IPU. 
\end{enumerate}

The rest of the paper is organized as follows: Section two defines the problem in more details, while section three reviews related work in this area. In section four, benchmark results are expressed. Finally, a conclusion is discussed in section five.



\section{Background}

\subsection{Intelligent Processing Units (IPUs)}
The Graphcore Intelligence Processing Unit (IPU) is a highly parallel processor and  has been especially designed for a large amount of parallelism. This makes it 
particularly well suited for both dense and sparse workloads in a wide range of applications including machine learning and graph processing algorithms.

In this work, a M2000 IPU Pod-4 system containing four GC200 is used. This is the second IPU generation, improving overall memory capacity and the number of cores compared to the first generation GC2 processor. During this work, the third generation IPU, Bow, based on wafer-on-wafer-technology has been released.

The graphcore IPU consists of two basic building blocks: IPU-Tiles and the IPU-Exchange. Each IPU-Tile consists of an IPU-Core and In-Processor-Memory. In contrast to a GPU, where each execution unit is only able to execute in a SIMD fashion, each IPU-Core is able to schedule six threads in a time-sliced fashion. This means that it is capable of executing in a MIMD fashion. The In-Processor-Memory of an IPU contains a small amount of SRAM memory.

\subsection{Poplar \& Poplibs}
Apart from the common ML frameworks such as Tensorflow and PyTorch, Graphcore is also providing low-level programming features with its Poplar C++-framework.

\begin{figure}[htbp]
	\centerline{\includegraphics[width=90mm]{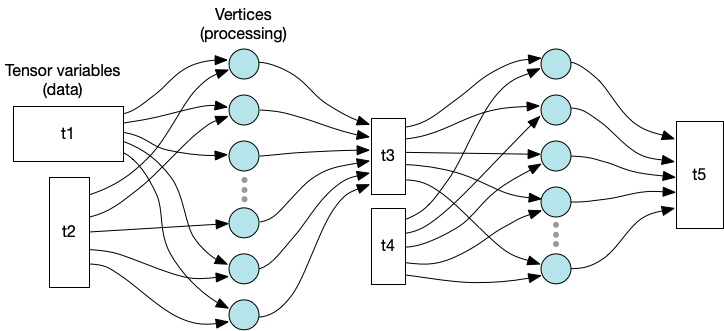}}
	\caption{Poplar Computational Dataflow Graph \cite{b8}}
	\label{fig: graph}
\end{figure}

IPU-Programs are represented by a computational dataflow graph as depicted in Fig. \ref{fig: graph}. In this graph, computation is represented as nodes (Vertices), data as Tensors and the flow of these as Edges. Those Vertices consist of Vertex code and connected data as Tensors. Each Vertex can be mapped to individual Tiles and be executed independently.

This graph can either be created via Poplar API or Poplibs libraries. In Poplar, the Vertex code executed on each Tile and the mapping of Tensors and Vertices to Tiles is done explicitly. 
The PopLibs libraries provide application-level functions that can be used in programs for the IPU. These offer functionality to implicitly add common operations like a MM to the computational graph. Additionally, the mapping of vertices and copying of data is also part of PopLibs.

\subsection{Memory Considerations}
\begin{figure}[htbp]
	\centerline{\includegraphics[width=90mm]{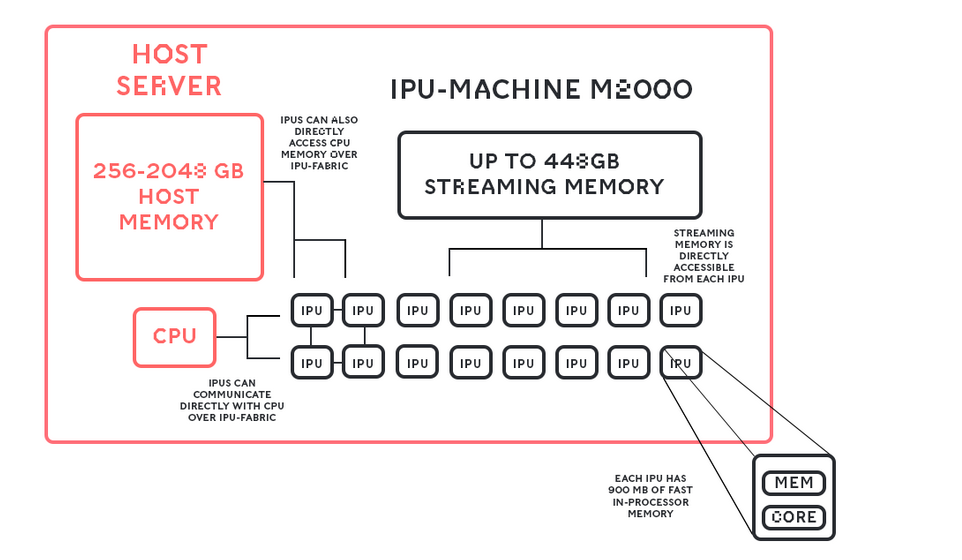}}
	\caption{IPU Memory Hierarchy \cite{RemoteBuffer}}
	\label{fig: memory_hierarchy}
\end{figure}

An IPU-Tile consists of an IPU-Core and In-Processor Memory.
In contrast to a GPU, the only memory directly accessible by each processing unit is the In-Processor memory local to its tile. For a GPU, several levels of memory can be accessed by each processing unit. This means that all data required for a computational step  must reside in the In-Processor Memory of each tile for an IPU.

As soon as the data is available on the IPU it can be processed. This work deals only with the execution of the matrix multiply and views the data as already available. 

While PopLin is the IPU's natural counterpart to an NVIDIA GPU's cuBLAS, it is currently lacking support for multiple IPUs, limiting associated experimentation to a single IPU.

\subsection{Workload Selection}

In HPC, AI, and ML applications, matrix multiplication (MM) is such a common workload that its performance is often used as an indicator of an architecture's overall efficiency. This workload is included in almost all benchmarks such as SPEC \cite{SPEC}, NVIDIA CUDA Sample SDK \cite{NVIDIAcuda}, Rodinia Benchmark Suite \cite{Rodinia}.

In \cite{IPUDiss} strong limitations in maximum problem size for MM on a GC2 IPU are mentioned. The peak performance achieved is 18.9 TFlops/s with a matrix dimensions of 2944 x 2944. With squared matrices, this translates to 104MB which equals 35\% of available In-Processor Memory of one IPU. The performance is 60.7\% of theoretical peak performance.

First experiments verified this behavior for the GC200 IPU with a maximum performance of 43.3 TFlops/s, 69.3\% of theoretical peak performance. The matrix dimensions are 3584 x 3584 (154MB) which equals 17\% of available In-Processor Memory. This is also the largest problem size fitting onto the IPU.

The limits mentioned in \cite{IPUDiss} refer to squared matrix multiply. As the underlying memory architecture and execution differ between GPU and IPU, experiments examining these differences are of great interest. A good candidate for that is the skewed MM in which the two dimensions of A are varied.

Therefore, this work examines the behavior for MM on the IPU. Key focus for this work is two-fold:
\begin{enumerate}
	\item Performance evaluation of squared and skewed matrix multiplication impolementation on the IPU.
	\item Comparison of skewed MM on IPU with GPU.
\end{enumerate}
In all experiments, the problem of matrix multiplication is considered as $A[m,n] \times B[n,k] = C[m,k]$ with single-precision floating point values.

\subsection{How GPU implementations differ from IPU}
For IPUs, the implementations differ to those for a GPU.
This is mainly due to the underlying memory architecture and the execution model. 
In contrast to a GPU executing in a single instruction multiple threads (SIMT) fashion, each IPU-Core can execute programs independently (MIMD). An IPU-Core is able to schedule six threads in a round-robin scheme. 

As the IPU itself does not have a memory hierarchy and each IPU-Core can only access its own In-Processor-Memory, data has to be moved between IPU-Tiles between the computational phases. In Fig. \ref{fig: memory_hierarchy}, types of memory available to the IPU are shown. There are two main ways to exchange data with the IPU: Either via host memory residing on the host server or via Remote Buffers residing in Streaming Memory.  

Therefore, the IPU relies on the Bulk-Synchronous Parallel model introduced by \cite{valiant} in which the processor executes in supersteps. The three phases for an IPU are:

\begin{figure}[htbp]
	\centerline{\includegraphics[width=70mm]{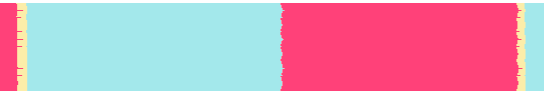}}
	\caption{Different phases of a program represented in the profiler: red: BSP Superstep compute, blue: synchronization, yellow: data exchange}
	\label{fig:exec}
\end{figure}

1. local tile compute: Fig. \ref{fig:exec}, red

2. global cross-tile synchronization: Fig. \ref{fig:exec}, yellow 

3. data exchange for the IPU: Fig. \ref{fig:exec}, blue \cite{b8}

\section{Literature Review}
IPUs are a acomparatively new hardware platform for running massively parallel programs. This results is completely new research paths, and hence, there are currently few works or experiences published in research community. Several reseach focus on AI applications of this hardware, however, there exists some research that target general HPC usage of IPUs.

Nasari et al.  \cite{ipu_ML_bench1} provide a comparison of GPUs and IPUs by running training benchmarks of common AI/ML models. They examine that some training parameters like batch size has important effects on the performance because of the different memoy yub-system in these two architectures. More specifically, IPUs run efficiently with smaller batch sizes, while GPUs benefit from large batch sizes to extract sufficient parallelism in neural network processing.

Most current papers like \cite{Conf-ML}, \cite{Dyn_Sparse}, \cite{Qwant}, and \cite{literature3} examine the IPU in its advertised use. They describe the use of IPUs for accelerating AI/ML workloads. Since we are looking at IPUs in a general HPC context, using the framework level libraries is not viable in our case. 

In the scope of graph processing, which are samples of irregular pattern parallel algorithms, there exists some works like  \cite{ipu_graph1} to implement distributed algorithms with sparse structure on early generations of IPUs. Luk et al.  \cite{Luk1} also provide a deep analysis of BFS implementations on IPUs to examine hard-to-predict memory accesses by BFS or BFS-based algorithms.

Luow and McIntosh-Smith \cite{IPU_HPC} use the GC2 IPU for stencil computations on structured grids and characterize the performance with STREAM memory benchmark results and a roofline model. They inspected Poplar, the low-level programming framework on IPUs and reported sufficient programmability to implement these HPC problems, and achieve performance compared to that of modern GPUs.

In \cite{IPUDiss}, the authors dissect an IPU-Pod16 system consisting of GC2 IPUs in terms of computational performance and peak bandwidth (inter and intra IPU). They also reported that in single precision, one IPU processor offers almost twice as much single-precision theoretical throughput as one V100 GPU: 31.1 vs. 15.7 TFlops/s.

However, as skewed MM is not included in both papers, we include this in our considerations. 
Furthermore, since memory is a scarce resource on the IPU, one has to gain a deeper understanding how Tensor data translates to memory usage.

\section{Methodology and Implementation}
Several experiments need to be performed to expand the insights of prior work. Testing the general performance of MM is the first objective to see where we stand with regards to the 62.5 TFlops/s of theoretical peak performance. Since GPUs are successfully used to accelerate MM computations, GPU performance is used as a baseline.

First, squared MM is performed on GPU and IPU which will serve as a baseline. Then, skewed MM is performed with different input matrix aspect ratios and analyzed using PopVision Graph Analyser and NSight Compute. For the implementation, we use the libraries cuBLAS for the GPU and Poplin for the IPU.

\subsection{Hardware details}
The system used in this work is a M2000 IPU-Machine connected via 100 GbE to the host server. The M2000 consists of four GC200 IPUs with 512 Gbps of inter-IPU communication bandwidth provided by IPU-Link \cite{b9}. For the MM comparison with the GPU, an NVIDIA A30 is used since they are very close in terms of clock frequency and power consumption. the table to compare our IPU and GPU. A comparison of the two architectures is shown in \ref{table:Arch_diff}.

\begin{table}
	\centering
	\caption{Comparison of IPU GC200 and GPU A30}
	\begin{tabular}{|l|c|c|c} 
		\hline
		Chip & GC200 & A30 \\
		\hline
		Number of cores     & 1472 	& 3584 \\
		Number of threads   & 8832 & 229,376 \\
		Total SRAM          & 918 MB & 10.75 MB \\
		Total DRAM memory    & 256 GB & 24 GB \\
		DRAM bandwidth         & 20 GB/s & 933 GB/s \\
		Clock frequency     & 1.33 GHz & 1.44 GHz \\
		FP32 peak compute     & 62.5 TFlops/s    & 10.3 TFlops/s \\
		Power Consumption      & 150 W & 165 W \\
		Inter-Chip-Bandwidth & 350 GB/s & 200 GB/s \\
		
		\hline
	\end{tabular}
	\label{table:Arch_diff}
\end{table}

\subsection{Profiling procedure}
For profiling the IPU, we used the PopVision Graph Analyser \cite{popvision} provided by graphcore. For the GPU, we used NSight Compute provided by Nvidia. Since the hardware differs for IPU and GPU, few comparable metrics are present. One important metric for the analysis is the execution time for the calculation of the matrix multiplication excluding data movement. Another important metric is the Tile Utilisation (IPU) and the Achieved Occupancy (GPU). Since the IPU only has one type of memory, further analysis on how the memory is used is necessary. PopVision offers this as well.

\section{Results and Discussion}

In this section, the results of different implementations of squared and skewed matrix multiplication on both the IPU and the GPU are shown and discussed.

\subsection{Results}
\begin{figure}[htbp]
	\centerline{\includegraphics[width=100mm]{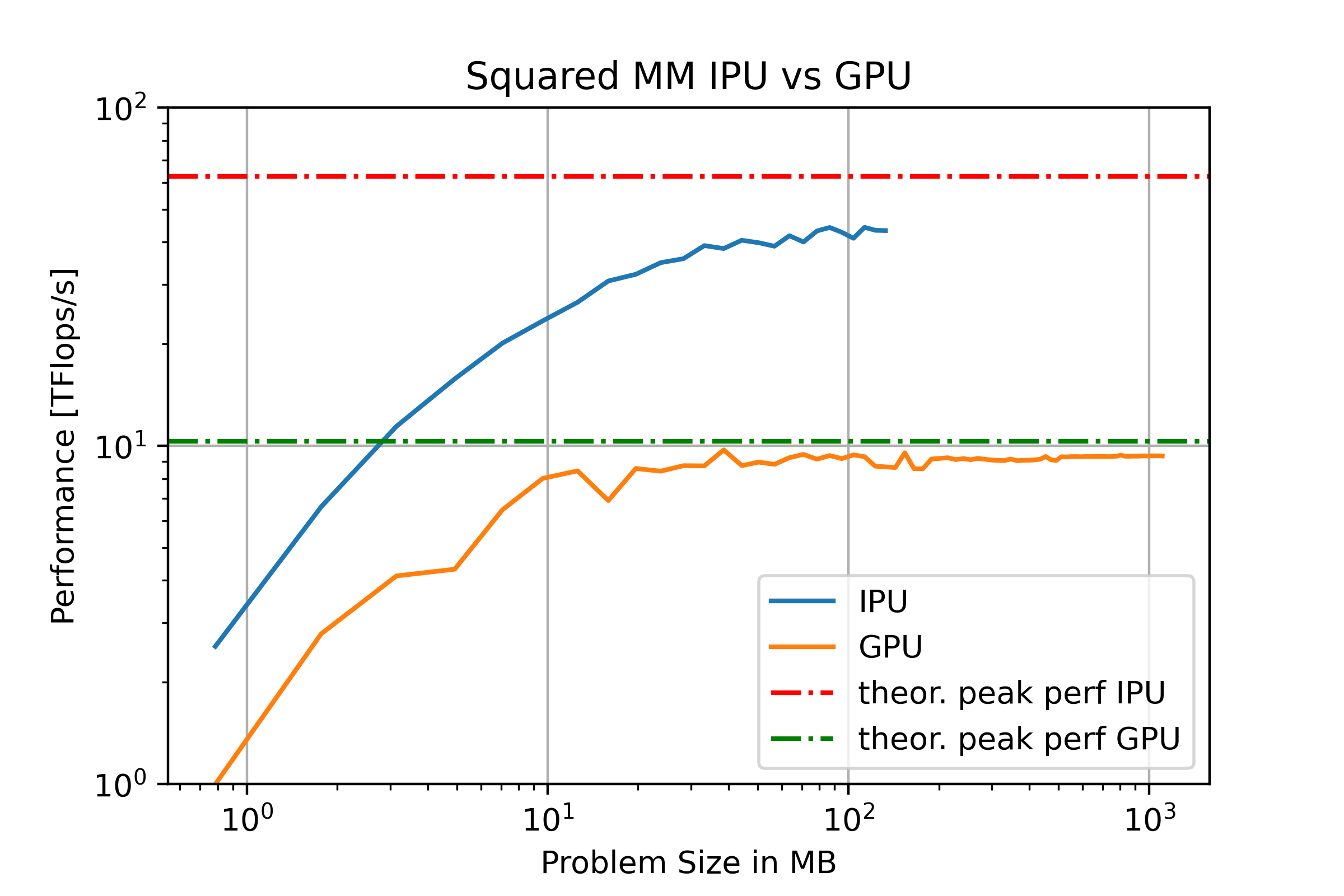}}
	\caption{Performance of squared MM}
	\label{fig: GPU_perf_new}
\end{figure}

Fig. \ref{fig: GPU_perf_new} shows the performance of a squared MM on the IPU and the GPU for different problem sizes. The horizontal lines represent the theoretical peak performance of 62.5 TFlops/s and 10.3 TFlops/s for IPU and GPU, respectively. While the GPU almost achieves theoretical peak performance with 9.7 TFlops/s, the IPU falls short with 44.2 TFlops/s. The latter performance number was verified by the IPU's manufacturer. Despite this fact, the IPU still surpasses the GPU's performance as long as the matrices can fit into the IPU memory. A known drawback of the IPU is the lack of memory, as can be seen, the GPU can handle larger data sizes.

\begin{figure}[htbp]
	\centerline{\includegraphics[width=100mm]{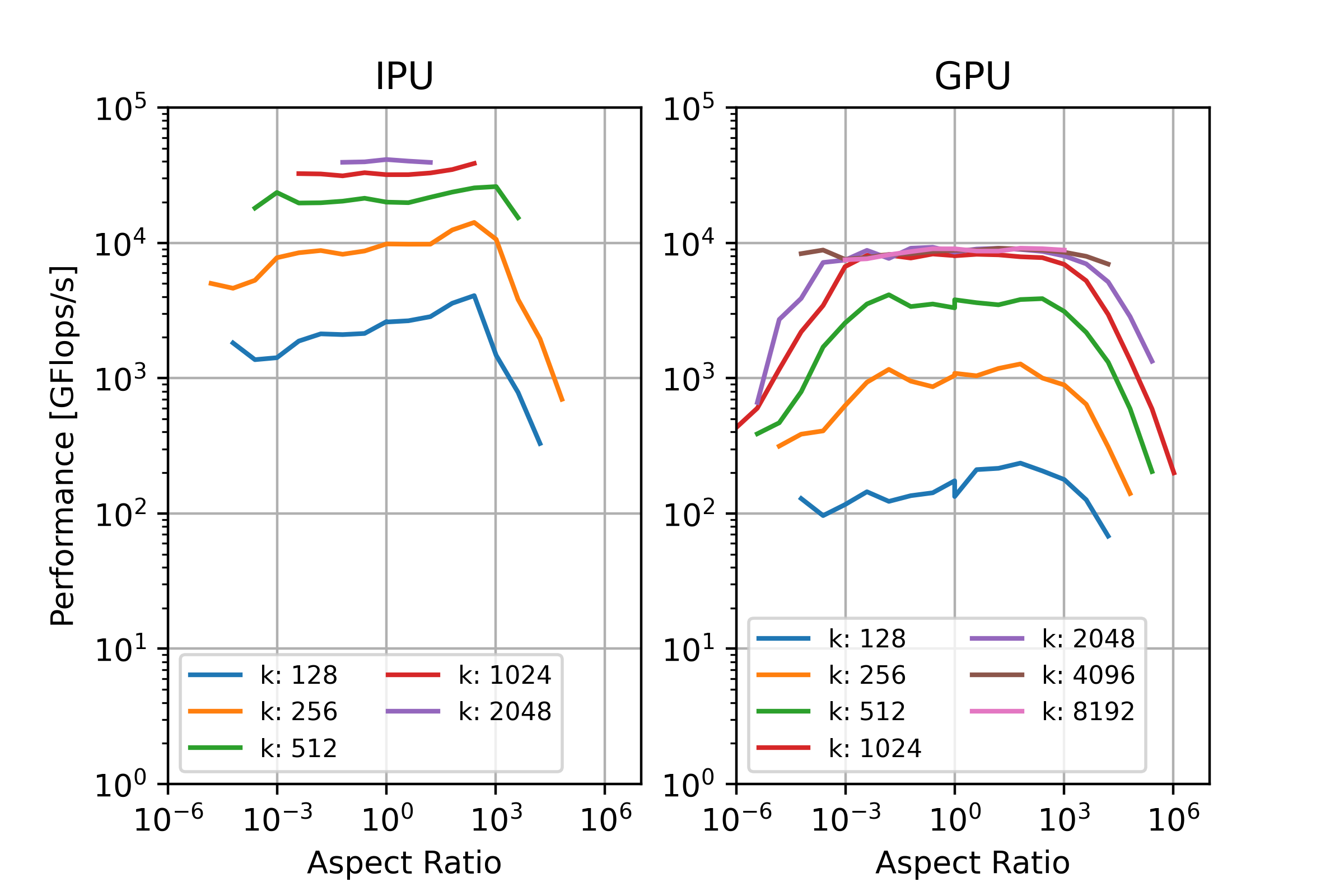}}
	\caption{Performance of skewed MM}
	\label{fig: comp_new}
\end{figure}

Fig. \ref{fig: comp_new} shows the results of MMs on the IPU on the left and the GPU on the right. In this experiment, different aspect ratios are used to see whether the IPU can deal with skewed MM and squared MM better than the GPU or not. The matrices multiplied have the dimensions of $m \times n$ and $n \times k$. Specifically, $k$ is varied for the GPU and IPU to keep the aspect ratios but vary the data size.

For the GPU, high aspect ratios to both sides result in significantly lower performance. For the IPU, on the left side of the graph (left-skewed MM), a performance decrease can be observed as well as on the right side (right-skewed MM). However,  in contrast to the GPU, the performance drops are not as symmetrical on the IPU. For the right-skewed MM, the performance drop is much more severe. 
Overally, the IPU seems to have less issues and therefore less performance loss when dealing with skewed MM.

In order to analyze the performance drop behavior when moving towards the right-skewed aspect ratios on the IPU, statistics and metrics collected by the PopVision tool are evaluated. Evidences show a large difference between the number of vertices for the three experiment types: left-skewed, squared, and right-skewed. The corresponding numbers for a given k are 5542, 5762, and 31743, respectively. This huge difference makes it much unpredictable to estimate the behavior of the compiler and the runtime and finally results in the shown performance drop in right-skewed MM.

\subsection{Discussion}

After evaluating the results, there can be important findings mentioned as follows:

\textit{Finding 1}: While a GPU was already trading less memory capacity for more compute power (in comparison to a CPU), an IPU drives this even further. This is mainly because of both the memory and compute architecture of an IPU. Especially since the capacity of 900MB per tile In-Processor memory is considerable much less than the main memory in a GPU. Bringing higher parallelism costs in less total available memory, while keeping enough data close to the processing units.

Looking at real-world applications, skewed matrices are dominant in the field of AI and ML. When performing skewed MM, the IPU surpasses the GPU in terms of performance for all aspect ratios as long as they fit into the IPU's In-Processor memory. However, this has to be examined for problem sizes which do not fit in In-Processor memory. In this case, the streaming memory can be an option that should be examined in both accessibility and efficiency aspects.

\textit{Finding 2}: Results illustrate that the generated number of vertices by the compiler play a significantly important role for different types of matrix multiplcation execution and therefore results in different performance.

\textit{Finding 3}: IPUs show more robust behavior when input matrices are skewed to represent more practical experiments. The distributed memory and computation nature of this architecture provides desired robustness when dealing with even irregular parallel programs.

\section{Conclusion and Future Work}
During recent years, Graphcore IPUs has been employed for scaling the learning problems and other parallel scientific programs. Matrix multiplication has been always considered as a standard benchmark to evaluate different hardware platforms. In this paper, a thorough evaluation of running different variations of matrix mutliplication on one IPU is performed to examine the behavior of this hardware in different cases. 

Matrix multiplcation, as a widely used practical benchmark is used in different cases including the squared and skewed versions to evaluate the performance of IPUs. The results represent that IPUs can bring higher and more robust performance compared to similar GPUs specially when dealing with skewed matrices.

The main point to keep in mind is that on the IPU, memory is always the bottleneck. Several challenges were discussed, like the implicit mapping of memory by PopLibs or the memory footprint generated when mapping tensors to tiles manually. It has to be mentioned, that the IPU host contains memory as well, which can be used for streaming access, offering several IPUs the possibility to overlap communication and computation by reading data from host memory simultaneously.

The remote memory can be used for remote buffers as well, which are treated like every other memory on the system and can be copied to and from similar to In-Processor memory. This however, needs to be handled explicitly. This will become a further point of investigation. Despite the possibility of lower level optimizations, there are options to generate combined load store operations for more than one operand.

Experiments show that specifying proper AMP plays a significant role and can drastically affect both achievable peak performance and maximum input size. Future research can also consider how this parameter can affect the utilization of the remote memory.

Future works can also include experiments on multiple IPUs since there will be promising scalability benfits when using multiple IIPUs. In this regards, there might be improvements in either the maximum processable matrices or the perforamnce. However, there will be some limitations when working with Poplar SDK on multiple IPUs.

\section{Conclusion and Future Work}
Graphcore IPUs have been introduced to scale learning problems and other scientific applications. 
Matrix multiplication has been always considered as a standard benchmark to evaluate different hardware platforms. In this paper, a thorough evaluation of skewed matrix multiplication on IPU is performed to examine the behavior of this hardware for real-world workloads. 

Experiments show that the IPU behaves smoothly when dealing with different aspect ratios, while there are also some special different behavior in some high-skewed matrices. The main point to keep in mind is that on the IPU, memory is always the bottleneck. The maximum problem size that can be handled by the IPU is much less than the GPU.


Future works will include experiments with streaming access which involves IPU host memory for an increased overall memory capacity, remote memory which similarly extends memory capacity by relying on other IPU's memory, and scaling to multiple IPUs in general. In this regard, there are hopefully substantial improvements in either the maximum processable matrices size or the computational performance, thus maintaining excellent absolute performance and also performance robustness, while extending the application space.



\bibliography{IPUPaper}

\end{document}